# Scalable Intelligence-Enabled Networking with Traffic Engineering in 5G Scenarios for Future Audio-Visual-Tactile Internet

Yiqiang SHENG

School of Electronic, Electrical and Communication Engineering, University of Chinese Academy of Sciences,
No. 19(A), Yuquan Road, Shijingshan District, Beijing 100049, China
National Network New Media Engineering Research Center, Institute of Acoustics,
Chinese Academy of Sciences No. 21, North Fourth Ring Road, Haidian District, Beijing 100190, China
Corresponding email: shengyq@dsp.ac.cn

**ABSTRACT** In order to improve future network performance, this paper proposes scalable intelligence-enabled networking (SIEN) with eliminating traffic redundancy for audio-visual-tactile Internet in 5G scenarios such as enhanced mobile broadband, ultra-reliable and low latency communication, and massive machine-type communication. The SIEN consists of an intelligent management plane (ImP), an intelligence-enabled plane (IeP), a control plane and a user plane. For the ImP, the containers with decision execution are constructed by a novel graph algorithm to organize objects such as network elements and resource partitions. For the IeP, a novel learning system is designed with decision making using a congruity function for generalization and personalization in the presence of imbalanced, conflicting and partial data. For the control plane, a scheme of identifier-locator mapping is designed by referring to information-centric networking and software-defined networking. For the user plane, the registrations, requests and data are forwarded to implement the SIEN and test its performance. The evaluation shows the SIEN outperforms four state-of-the-art techniques for redundant traffic reduction by up to 46.04% based on a mix of assumption, simulation and proof-of-concept implementation for audio-visual-tactile Internet multimedia service. To confirm the validity, the best case and the worst case for traffic offloading are tested with the data rate, the latency and the density. The evaluation only focused on the scalability issue, while the SIEN would be beneficial to improve more issues such as inter-domain security, ultra-low latency, on-demand mobility, multi-homing routing, and cross-layer feature incongruity.

**INDEX TERMS** Next generation networking, intelligence-enabled networking, 5G scenarios, information-centric networking, software-defined networking, traffic engineering, scalability.

## I. INTRODUCTION

Essentially, networks and networking exist in order to transfer entities, objects, energy or information as traffic from ingress nodes to egress nodes. Enhancing the network traffic performance such as the scalability [1] has been widely investigated as one of the fundamental topics of network engineering, which deals with the issues of network modeling, measurement, characterization, evaluation and optimization by applying scientific principles, emerging techniques and engineering strategies. However, it is urgent to consider higher performance for the fifth generation (5G) [2][3] mobile communications system and beyond it [4][5]. Typical 5G scenarios include enhanced mobile broadband (eMBB), ultra-reliable and low latency communications (URLLC), and massive machine type communications (mMTC). For example, the next generation networking such as tactile Internet [6] will meet the requirement of ultra-low latency such as shorter than 1 *ms* and ultra-high area traffic capacity such as 10 Mbps/m$^2$ or even higher at a wide range of moving speeds such as 500 km/h or even faster. That brings a big challenge to traffic engineering for academic research and industrial development.

    Currently, there are two of the most prevalent techniques for traffic offloading, peer-to-peer (P2P) [7] [8] with multiple hops and device-to-device (D2D) [9] [10] [11] with single hop, to implement large-scale content delivery [12]. To satisfy the requirements for offloading traffic and mitigating network congestion, P2P and D2D are used to exploit the idle resources such as storage space and bandwidth capability from peers and devices based on the collaboration of neighbors. In P2P or D2D, the content is allowed to be directly transmitted among personal computers or mobile devices. Furthermore, the integrity-oriented offloading [13] and the D2D-based information-centric networking (ICN) traffic offloading [14][15] has been introduced to deliver the resources for improving traffic and mitigating congestion, and many researchers proved the effectiveness of eliminating redundant traffic by ICN techniques.

    In the near future, we have to be targeting to satisfy more 5G requirements in terms of scalability for a massive number of objects, high security, low latency, mobility on demand, effective compatibility, efficient manageability, and so on. However, the current Internet architecture that IP addresses are used as the common type of identifiers and locators for mapping and

routing would not be easy to meet the above requirements, since the location-dependent IP architecture has its inherent limitations in supporting on-demand mobility, scalable routing, etc.

Furthermore, the 5G specifications and standardization are still lacking so far, and mostly focus on network function virtualization (NFV) and software defined networking (SDN) [16]. The use of ICN [17] and its related techniques in 5G is under-researched. The ICN architectures, such as CBCB, DONA, NDN, NetInf, PSIRP, PURSUIT, MobilityFirst, are identified as one of the most effective ways to overcome the limitations of IP-based Internet. The CBCB [17] pushed and pulled interest packages to publishing and subscription. In DONA [18], the P: L flat naming based on unified resource identifier (URI) was used as a tree-topology RH resolution system. In NDN [19], the hierarchical names using readable URIs are introduced to realize the aggregation and the compatibility with TCP/IP. In NetInf [19], a REX system was designed to realize naming and name resolution using the MDHT [20] technique. The naming based on SID and RID and the routing based on Bloom filtering were proposed in PSIRP/PURSUIT [21]. In MobilityFirst, a global unified identifier was designed to satisfy the security and mobility requirements. The ICN addressed the issue of mobility in IP by allowing applications to bind to identifiers as names, where ICN manages the identifier-locator mapping in the network. The identifier could be any object such as mobile devices, data, service or a piece of content, while the locator could be any network addresses. Each identifier could be associated with at least one network address. The separation between identifiers and locators makes it possible to be topology-independent for publishers and subscribers.

In future, it will be possible to incorporate the advantages of the paradigms for intelligent SDN [22], NFV and ICN for processing a huge volume of data and handling a number of heterogeneous devices. For example, M. Zorzi *et al.* [23] advocated a learning model as the key building block to extract context representations for COBANETS. The concept of intelligent networks paved a way to the avenues that intersect multiple research domains.

However, with the increase of terascale ($10^{12}$), petascale ($10^{15}$) or even exascale ($10^{18}$) objects such as data, instead of the existing ten-billion-level ($10^{10}$) devices, the current networks suffer from the issue to deal with scalability due to the cognitive complexity of automaticity, personalization, and generalization, since it is inefficient for complete data manual processing to handle the large-scale network data in practice. The cognitive complexity means the gaps between how the human being determines relevance and calculates probability versus how computers or artificial intelligence determine similar concepts. The automaticity means the ability to handle data without manual processing, allowing it to become an automatic response. The personalization means tailoring a response, a service or a product to accommodate specific individuals, sometimes tied to some groups or segments of individuals, to improve customer satisfaction, quality of service (QoS), or quality of experience (QoE), etc. The generalization is the process of identifying the parts of a whole or learning the complete information through the partial data.

By focusing on the scalability issue this time, the main contributions of this paper are as follows. (1) Scalable intelligence-enabled networking (SIEN) is proposed in order to improve future network performance. An intelligence-enabled plane (IeP) is provided by decoupling the function of learning and decision making from the management and control planes. On the IeP, a novel congruity learning system is designed with generalization and personalization. A novel weighted graph algorithm for container generation is designed to efficiently manage large-scale distributed nodes and resource partitions by taking advantage of local affinity and executing the decisions to satisfy the 5G requirements. (2) The SIEN is realized using ICN schemes such as identifier-locator mapping, routing and forwarding to test the performance in typical 5G scenarios such as eMBB, URLLC and mMTC. The evaluation showed up to 46.04% improvement for the scalable performance of 5G traffic offloading with respect to the data rate, the latency and the density in comparison to four of the state-of-the-art techniques. Furthermore, the best case and the worst case (at least 5.56%) are tested to confirm the validity and the stability of eliminating traffic redundancy with a mix of assumption, simulation and proof-of-concept implementation to improve 5G traffic offloading for future audio-visual-tactile Internet.

The rest of this paper is organized as follows. Section II overviews the architecture of scalable intelligence-enabled networking. Section III introduces the intelligent management plane. Section IV describes the intelligence-enabled plane. Section V presents the control and user planes. Section VI shows the evaluation. Section VII provides related work. Section VIII is the analysis and discussion. Section IX concludes this research.

## II. ARCHITECTURE

This section shows the architecture of scalable intelligence-enabled networking (SIEN) that includes an intelligent management plane, an intelligence-enabled plane, a control plane and a user plane. The section also shows the relations among different planes, and how the planes communicate each other.

Intelligent management plane (ImP): A container means a set, cluster or class of objects that may include information, predictive resource partitions [24], and network elements such as servers, switches, personal computers and mobile devices. The containers would be constructed to execute the decisions for managing objects by the graph algorithms, which are designed to organize distributed nodes and resource partitions by taking advantage of local affinity. The containers can be also responsible for monitoring the network elements, collecting the data, and processing the network states, events, and services to improve the network performance in the long term.

Intelligence-enabled plane (IeP): The learning algorithms are designed with generalization and personalization to deal with the cognitive complexity for predicting situation, making judgments, eliminating network traffic redundancy, discovering issues, fixing detected problems, improving automaticity, and meeting the uncertainties of environment. The IeP integrates the critical principles of the knowledge plane [25], the inference plane [26] and the heterogeneous networks [27] based on artificial intelligence responsible for representation, learning, reasoning, mining, complex event processing, and situation awareness to optimize the networks in a global view.

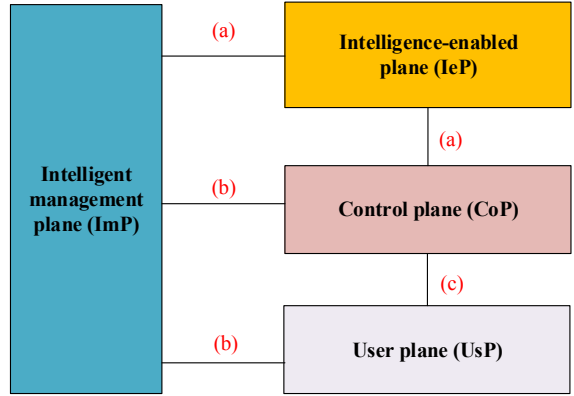

NOTE: (a) Learning and Decision Making (LDM); (b) Containerization and Decision Execution (CDE); (c) Identifier-Locator Mapping (ILM).

**FIGURE 1.** The architecture of scalable intelligence-enabled networking includes an intelligent management plane, an intelligence-enabled plane, a control plane and a user plane.

Control plane (CoP): By referring to the principles of software-defined networking (SDN) and information-centric networking (ICN), a container-based scheme on at least one middle level is used to implement the naming, identifier-locator mapping, and routing. The controllers on the CoP exchange operational states to update the processing rules of the following user plane, as shown in FIGURE 1.

User plane (UsP): The registrations, requests and data are forwarded between network elements such as switches by referring to the principles of SDN and ICN. The registrations are available for naming and name resolution. The requests are available for name resolutions and data. The UsP should operate at packet-level time scales. The network elements on the UsP are responsible for storing, forwarding and processing data packets. That is to say, the UsP is the part of a network, which bears the user traffic such as the forwarding data that the network exists to carry.

Learning and decision making (LDM): The LDM is designed for making judgments, decisions, inferences, and predictions in the presence of imbalanced, conflicting and partial data. It monitors the network elements intelligently for connecting IeP, ImP and CoP. The LDM supports the traffic classification, the prefetching replacement strategy, and the predictions of distance, quality, cost and utilization for storage, bandwidth, and computation. As the SDN decouples the controllers from the user plane, the SIEN decouples the LDMs from the ImP and the control plane, and provides the IeP. Each element on the ImP has the LDM interface agent, and the corresponding element on the IeP or the CoP has the LDM interface driver. The IeP communicates with the ImP and the CoP to gather the data and return the learning results and decisions though the LDMs.

Containerization and decision execution (CDE): The CDE is designed by executing decisions based on containerization for achieving the congruity between the local events and the nonlocal situation. All measurable or predictable network elements can be organized in a sequence of hierarchical containers with more than three levels. The IeP would be on the top level. The control plane could be on at least one middle level. The UsP would be on the bottom level. Each element on the ImP can have a CDE interface agent, and the corresponding element on the CoP or UsP can have a CDE interface driver. The ImP communicates with the CoP and the UsP though the container to execute the decision from the IeP and to measure the network traffic, service, configuration, operation, etc.

Identifier-locator mapping (ILM): The ILM is designed as the foundation of routing to handle large-scale objects such as data, contents, services, devices, sensors, and vehicles. It can be realized by SDN based on ICN which aims to shift send-receive model towards publish-subscribe model. It is recommended to support naming based on self-certification, block chain or public key infrastructure for security. The ILM supports dynamic bindings between identifiers and locators, instead of static IP addresses, and multipath routing, since the locations of services and users are likely to be changed frequently in future Internet.

The locators can be a kind of identifiers, so the ILM would be a mapping among identifiers corresponding to objects. Each element on the UsP can have an ILM interface agent, and the corresponding element on the CoP can have an ILM interface driver. Though the ILM, the CoP communicates with the UsP by registering objects as the identifiers, i.e. naming registration, binding at least one locator with each identifier, i.e. resolution registration, and resolving the locators, i.e. network addresses or IPs.

## III. INTELLIGENT MANAGEMENT PLANE

This section describes how the objects such as network elements are organized by CDE as a sequence of containers on the ImP.

As mentioned in section II, a container can be a set of network elements or resource partitions. In practice, the containers can be constructed for managing nodes and providing services by making the best of information-centric networking (ICN), peer-to-peer (P2P) and device-to-device (D2D) techniques to satisfy the on-demand mobility of publishers and subscribers. The containers integrated all kinds of distributed nodes such as ICN servers, switches, personal computers and mobile devices.

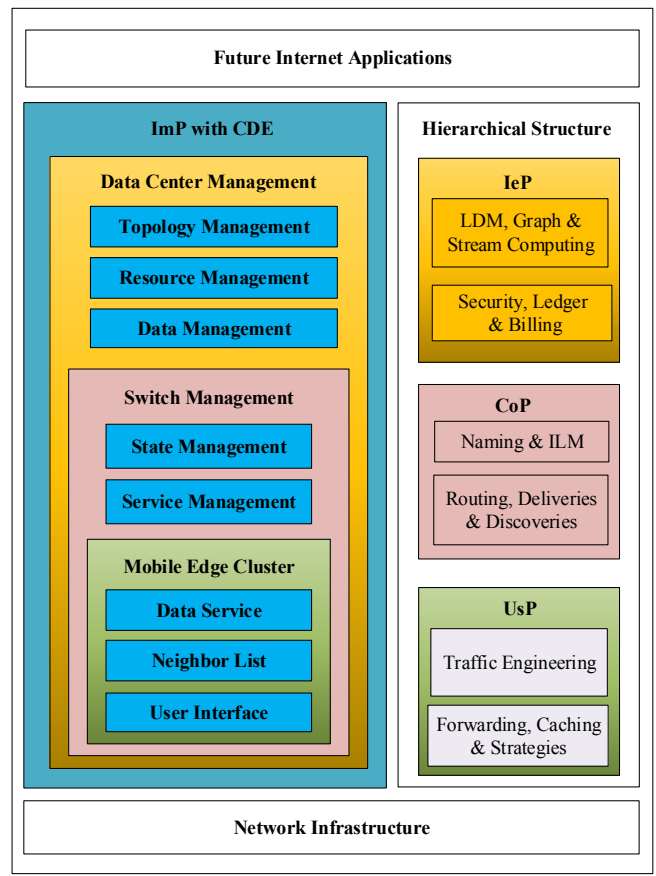

**FIGURE 2.** The SIEN deployed by multi-level containerization with hierarchical structure between future Internet applications and network infrastructure.

As shown in FIGURE 2, the SIEN can be deployed with hierarchical structure between future Internet applications and network infrastructure. In the SIEN, the ImP with the function of CDE. The IeP is with the function of LDM, graph/stream computing, security, ledger and billing. The CoP is with the functions of naming, ILM, routing, deliveries and discoveries. The UsP is with the functions of traffic engineering, forwarding, caching and strategies. A typical system of multi-level container consists of a data center management subsystem on the high level, a switch management subsystem on the middle levels, and a cluster of mobile edge nodes on the low level.

The data center management subsystem on the high-level container manages switches, distributed nodes and resource partitions for storing and processing information objects. The subsystem is divided into three functional modules: topology management, resource management and data management. The topology management is designed for gaining the node-link states, updating the statistical information, and managing the topological structure. The resource management is designed for collaborating with the data publishers, gaining the data sources and prefetching data copies. The data management is designed for preprocessing the history of user behaviors and network measurement.

The switch management subsystem on the middle-level container takes charge of exchanging the information with the data center management subsystem and directly managing the clusters of mobile edge nodes. It consists of two functional modules: state management and service management. The state management module is designed for detecting the states of all end-user nodes and uploading the information to the data center management subsystem. The service management module is designed for delivering the data copies to the clusters of mobile edge nodes.

The cluster of mobile edge nodes on the low-level container consists of three functional modules: data service, neighbor list, and user interface. The data service is designed for searching an object at the local storage and judging whether the request is forwarded to higher levels in case of no existence of the requested object copies. The neighbor list is designed for maintaining the neighbors according to the measured or learned distances, where the distance is a numerical measurement or prediction of how far apart objects are. The user interface is designed for getting the requests, providing the data services and collecting the history of user behaviors.

Based on the measured or learned distances, we get a weighted graph $G(V, E, W)$ to construct the hierarchical containers, where $V$ is a set of network elements or resource partitions, $E$ is a set of connected edges between nodes or partitions, and $W$ is a set of weights with measured or learned distance. The containers are multiple levels as the following. A high-level container marked as $C_{i+2}$ consists of middle-level containers such as $\{C_{i+1(1)}, C_{i+1(2)}, \ldots\}$ or low-level containers such as $\{C_{i(1)}, C_{i(2)}, \ldots\}$, where $1 \leq i \leq I$ and $I$ is a constant. A sequence of targets for distances $\{T_i\}$ is required for containerization. The sequence means an enumerated collection of at least two targets in which repetitions are not allowed. The targets mean the maximum distances required in at least two scenarios for applications. A set of containers $\{C_{ik}\}$ can be constructed for each $T_i$, where $1 \leq k \leq K_i$ and $K_i$ is the number of containers for a given $T_i$ according to the following algorithm.

| Novel Algorithm 1: Containerization |
|---|

Input:
    $G(V, E, W)$: A weighted graph;
    $\{T_i\}$: a sequence of targets;
Output:
    $\{C_{ik}\}$: a set of containers for each $T_i$;
Parameters:
    $k$: a positive integer, $1 \leq k \leq K_i$
    $K_i$: the number of containers for a given $T_i$;
    $W = \{w_{ij}\}$: A set of weights based on distances;
    $w_{ij}$: A weight between nodes or partitions $v_i$ and $v_j$;
    $\{G_m(V_m, E_m, W_m)\}$: A set of sub-graphs of $G$;

Get the data of $G$ and $\{T_i\}$;
Initiate $\{G_m\}$ and $\{C_{ik}\}$;
For $i$ from $I$ to $0$
  $k = 1$;
  For each $w_n$
    If $w_n < T_i$ then
      $w_n = 0$;
      Update $\{G_m(V_m, E_m, W_m)\}$;
    End if
  For $m$ from 1 to $M$
    If $G_m$ is an isolated node, then
      $C_{ik}$ = a set of the isolated node;
      $k = k + 1$;
    Else
      $L$= the nodes set in $\{G_m\}$;
    Do
      Select a node $a_j$ in $L$;
      If $T_i$ is additively accumulative
        Search all paths smaller than $T_i$ (eg. latency) from $a_j$;
      Else if $T_i$ is minimally accumulative
        Search all paths larger than $T_i$ (eg. bandwidth) from $a_j$;
      Else

    Search all paths to hit $T_i$ from $a_j$;
   $C_{ik}$ = a set of all nodes on the paths in L;
   L = L - $C_{ik}$;
  While (L ≠ Φ)
Output $\{C_{ik}\}$ for any $T_i$, where $1 \leq k \leq K_i$.

## IV. INTELLIGENCE-ENABLED PLANE

This section introduces how a learning system works with LDM on the IeP based on a massive number of data collected from the history of Internet services.

 To make the subscript clear, $i$ is used for the measured scalar or vector, $j$ is used for the learned scalar or vector. The small notations $x$ and $y$ are scalars, while the capital notations $X$ and $Y$ are vectors. $D = \{X_i, Y_i\}$ denotes the set of all measured data. $D_p = \{X_{pi}, Y_{pi}\}$ denotes the set of the personal data for the personalized parameter learning, validating and testing. $D_g = \{X_{gi}, Y_{gi}\}$ denotes the set of the general data with a part of labels for the generalized parameter learning, validating and testing. The items of $X_i$ include the timestamp, the scenario type, the uplink traffic, the downlink traffic, the capability, the utilization, the object density, the latency, the storage space, the bandwidth state, the computational state, the neighbor list, the source, the destination, the protocol, the port, the payload, and the miscellaneous data. The items of $Y_i$ include the personal label, the general label, the distance label, the scalability label, the mobility label, the security label, the object state label, the prediction label, the classification label, the prefetching replacement label, the service quality label, and the cost label.

 The learned output vectors are marked as $\{Y_j\}$. The measured distances, which could be latencies using the ping tool or hops using the trace-route tool, are the graph weights for containerization on the ImP. The learned distances on the IeP are used to replace a part of the measured distances in a changing network environment. To take the layer-wise learning networks as an example, the set of parameters can be represented by theta. Let $\theta = \{\theta_h\} = \{(\theta_{h1}, \theta_{h2}, ..., \theta_{hi}, ..., \theta_{hn})\}$ be the set of parameters, where $h$ is an order number of layers with $0 \leq h \leq H+1$, $H$ is the number of hidden layers, $h = 0$ means the input layer, $h = H+1$ means the output layer, $n$ is the number of parameters in a given layer, and $\theta_{hi}$ includes a vector $C_i$ for connection and a value $b_i$ for bias in a given layer.

 The positive input vector is
$$X_i = (x_1, x_2, ..., x_{d_i})^T.$$
where $d_i \leq d_{max}$ is the in-degree of the $i^{th}$ neuron, and $d_{max}$ is a constant for the maximum degree of all neurons. The generalization term is defined by regularization [28].
$$R_q = \|\theta\|_q = \left(\sum_j |\theta_j|^q\right)^{1/q}.$$
where $\theta_j$ can be any parameter, $q$ is a positive integer, $q = 1$ is for the Manhattan norm and $q = 2$ is for the Euclidean norm.

 As shown in FIGURE 3, a congruity learning system consists of the dataset, the generalization with memory, a filter, the personalization with response, and a congruity function. The congruity learning means that two sparsely activated networks, which are designed by simulating left and right brains, are trained by a congruity function to automatically optimize the parameters. The filter is defined by the top-$k$ cosine similarity [29][30].
$$F_k = \frac{X_i^p \cdot X_i}{\|X_i^p\|_k \|X_i\|_k}.$$
where $X_i^p$ means a transposed input from personal sampling, $k$ is a positive integer. The positive output scalar is
$$y_j = f(C_j X_j + b_j).$$
where $f$ is the activation function, $b_j$ is the bias of the $j^{th}$ neuron, $C_j = (c_{1j}, c_{2j}, ..., c_{ij}, ...)$ is the vector for connection, $c_{ij}$ is the weight from neuron $i$ to neuron $j$, and $i = 1, 2, ..., d_j$. The negative input vector is
$$Y_i = (y_1, y_2, ..., y_{d_i})^T.$$
where $d_i$ is the in-degree of the $i^{th}$ neuron. The negative output scalar is
$$x_j = g(C_j Y_j + b_j).$$
where $g$ is the activation function in the negative direction, $b_j$ is the bias of the $j^{th}$ neuron, $C_j = (c_{1j}, c_{2j}, ..., c_{kj}, ...)$ is the vector of connection, $c_{kj}$ is the weight from neuron $k$ to neuron $j$, and $k = 1, 2, ..., d_i$.

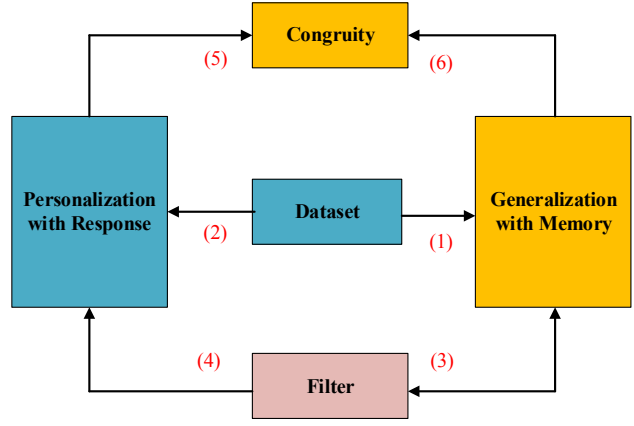

NOTE: (1) General Sampling; (2) Personal Sampling; (3) General Input; (4) Personal Input; (5) Personal Output; (6) General Output.

**FIGURE 3.** A congruity learning system consisting of the dataset, the generalization with memory, a filter, the personalization with response, and a congruity function.

The generalization is defined by a memory model with an error function using the general data.

$$E^g(q) = \sum_{y_i \in D^g} \lambda_g \|y - y_i\|^2 + \sum_{x_i \in D^g} \lambda_q R_q \|x - x_i\|^2.$$

where $\lambda_g$ is the weight for the training error based on the labelled general data and $\lambda_q$ is the weight for the $q$-th generalization term $R_q$.

The personalization is defined by a response model with an error function using the personal data.

$$E^p(\theta) = \sum_{y_i \in D^p} \lambda_p \|y - y_i\|^2 + \sum_{x_i \in D^p} \lambda_k F_k \|x - x_i\|^2.$$

where $\lambda_p$ is the weight for the training error based on the labelled personal data and $\lambda_k$ is the weight for the filter $F_k$.

The congruity function with its minimization is defined by

$$E(\theta) = (1 - \alpha)E^g(\theta) + \alpha E^p(\theta),$$
$$E^* = \min_\theta E(\theta),$$
$$\theta^* = \arg \min_\theta E(\theta).$$

where $E(\theta)$ is the congruity function, $0 \leq \alpha \leq 1$ is the weight of the congruity for generalization and personalization, $\theta^*$ is an optimized set of parameters, and $E^*$ is an optimized congruity function. To predict the distance, the learning system is trained and executed by the following algorithm.

---

Novel Algorithm 2: Congruity Learning

---

Input:
    $D_p$: The personal dataset;
    $D_g$: The general dataset;
Output:
    $\{Y_j\}$: The learning results;
Parameters:
    $\theta_h = (\theta_{h1}, ..., \theta_{hn})$: The parameters in the $h$-th layer;
    $n$: The number of parameters in a given layer;
    $h$: An order number, $0 \leq h \leq H+1$;
    $H$: The number of hidden layers;

Initiate all parameters such as $\theta_h$ ($0 \leq h \leq H+1$) as $\theta_i$;
For $h$ from 0 to $H+1$
    $d = n$;
    For all batches of data ($D_p$, $D_g$)

$\Delta\theta_j = -\lambda_j \partial E^p(\theta_j)/\partial \theta_j$;
$E_v^p = E^p(\theta_j)$ and $E_v^g = E^g(\theta_j)$;
$\theta_j = \theta_j + \Delta\theta_j$;
$E_h^p = E^p(\theta_j)$ and $E_h^g = E^g(\theta_j)$;
$\Delta E^p(\theta_j) = ABS(E_h^p - E_v^p)$;
$\Delta E^g(\theta_j) = ABS(E_h^g - E_v^g)$;
If $\Delta E^p(\theta_j) < \alpha \Delta E^g(\theta_j)$ then
    $\theta_j = 0$ and $d = d - 1$ with a user-defined probability distribution;
    If $d < d_{max}$ then
        $h = h + 1$;
        Break;
    End if
    End if
For all batches of data $(D_p, D_g)$
   Optimize $E(\theta)$ using error back propagation or heuristics;
Output $E^*$, $\theta^*$, and $\{Y_j\}$.

The global prefetching for Internet service would depend on the centrality of nodes, the popularity of objects and the path with the minimum distance. A prefetching strategy is used by delivering the top-$k$ popularity objects with local copies based on the graph with the distance weights, and the higher-centrality nodes have higher probability to be selected as the delivering node. If the popularity of an object is taken into account, then the probabilistic high-centrality prefetching is shown as the following.

$$p_{i,j} = \frac{nc_i fp_j}{\sum_{i,j} nc_i fp_j}.$$

where $p_{ij}$ is the selection probability to deliver an object with the popularity $fp_j$ to a node with the centrality $nc_i$ on the path with the minimum distance, $1 \leq i \leq I$, $1 \leq j \leq J$, $I$ is the total number of nodes to be selected and $J$ is the total number of objects to be delivered. The top-$k$ object popularity is based on the real statistics of service requests with an approximate Mandelbrot-Zipf distribution.

## V. CONTROL AND USER PLANES

This section describes how the ILM-based scheme works on the control and user planes to implement the SIEN.

The control plane includes a nonlocal ILM with naming service and distributed ILMs which deploys in nested containers as a tree topology, while the user plane includes forwarding elements, subscribers and publishers. Each ILM is with a table to map among human readable name (HRN), globally unique identifier (ID) and at least one network address (NA) such as IP in this paper, where an HRN could be described by uniform resource identifier (URI). The forwarding elements could be access points (APs), switches, routers, gateways, etc. The subscribers and publishers could be devices, servers, personal computers, sensors, etc.

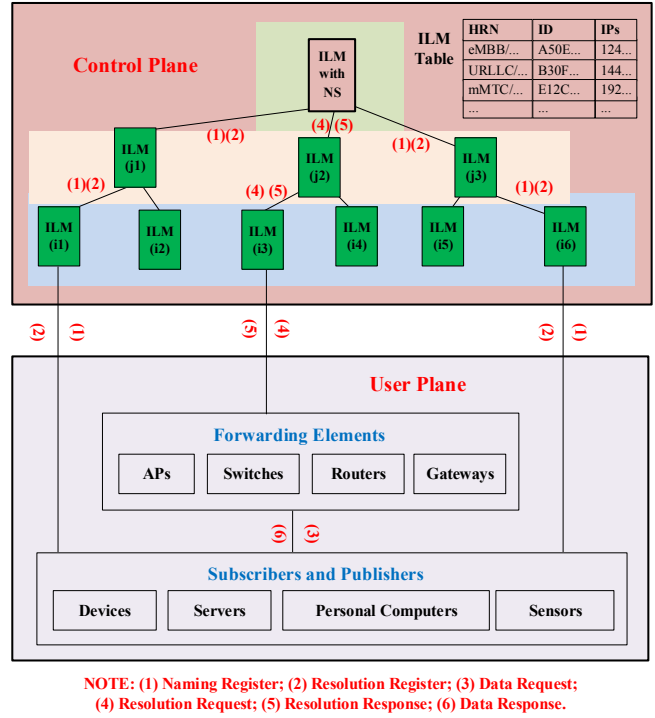

**FIGURE 4.** The block diagram of control and user planes with identifier-locator mappings (ILMs) using information-centric networking (ICN) and software-defined networking (SDN).

As shown in FIGURE 4, each subscriber or publisher can be registered for identifiers and locators (i.e. naming and resolution registration) via a local ILM, and the ILM would propagate the registration to higher-level ILMs until reaching the nonlocal ILM with naming service (NS). After the registration, a subscriber can send a request message with the ID of the requested object and its own ID to a forwarding element in order to get data. The forwarding element would ask a local ILM for a mapping between the destination ID and its IPs to get the IPs of the requested ID. The ILM would reply with a set of IPs as the resolution results. The forwarding element would select one of IPs, add it to the message and implement forwarding operation. It would be efficient for mobile destinations to obtain an updated list of IPs. The data forwarding procedure can be decoupled with the data request procedure by separately delivering the message.

For the relation between ILMs and containers, the ILM (j1) on the control plane corresponds to the container $C_{j(1)}$ on the ImP. If a publisher asks the ILM to register an ID with its IPs, the message first arrives a local ILM in the low-level container such as $C_{i(1)}$. Then, the message can be forwarded to a high-level container such as $C_{j(1)}$ or the higher-level container for the nonlocal ILM until returning the requested result. If the subscribers send a message including the ID of the requested object and its own ID to its local forwarding element, then the forwarding element first asks the local ILM such as $C_{i(3)}$ for a mapping between the destination ID and its IPs. Then, the message is forwarded to a high-level container such as $C_{j(2)}$ for replying with a set of IPs. When necessary, the message would be forwarded to the nonlocal ILM.

For the caching scheme, it supports on-path caching in the implicit way and allows to answer the requests for the same ID with the local copies. Based on passing messages at intermediate data forwarding elements, it allows subsequent requests for the same ID to be answered with the locally cached copies. Each time an object is cached off-path or replicated in the explicit way, the ILM should be informed of the change to update the corresponding ID entry with the additional IPs. Each forwarding element might implement its own policy on consulting with the ILM for additional cached copies.

In current Internet, IP addresses are used for the mapping of the host-centric or server-client models to identify a realm of the administrative domain. The domain name system (DNS) supports only static bindings, mostly using static IP addresses. The frequent change in a location of terminals or data objects would lead to failure in the end-to-end connection due to the use of IP address which is bound to the location.

In future Internet such as 5G, the ILM would be adapted to the dynamic bindings between identifiers and locators to support mobility on demand and satisfy the requirement of low latency. The identifiers for all objects are split with their dynamic IPs. By using a nonlocal ILM table, each object is assigned a globally unique ID via a naming service system that translates HRN to ID. Each ID is using a long identifier such as a hash string with 160 bits or even longer. Since the length ensures that the probability of a collision is small, the IDs can be randomly selected. Each ID can be mapped into one or more IPs via the ILM. If an object is available, all copies will have the same ID in multiple locations. It supports identifier-based delivery via data IDs or host-to-host communication via device IDs.

## VI. EVALUATION IN 5G SCENARIOS

In this section, the SIEN was evaluated by assumptions, simulation and proof-of-concept (PoC) implementation using information-centric networking (ICN) in 5G scenarios such as mMTC, eMBB, URLLC.

We assume that all globally unique long identifiers (IDs) for objects are with 160 bits fixed lengths as flat names and each ID binds less than five network addresses (NAs) with 32 bits using IPv4 addresses or 128 bits using IPv6 addresses for eMBB, URLLC, and the nonlocal domain of mMTC. As mentioned in section V, a global identifier-locator mapping (ILM) with naming service has been used to map from an HRN to ID. The additional information such as the type of service and its priority shares 32 bits lengths.

For the local domain of mMTC, it is not feasible to run long IDs due to the length of the packet header and the costly ILM considering the limited resources of mMTC devices. We assume all mMTC devices have less than 50MHz computation, less than 128 bytes communication capability, less than 50kB memory, and less than 300kB storage. Then, we assume all local short names for mMTC objects are with 8 bits fixed lengths, and only lightweight protocols such as Bluetooth low energy (BLE) could be used in the local domain of mMTC.

The number of IDs is set to be near $10^{10}$ based on the number of mobile devices, and the number can be increased to $10^{15}$ or even scaled up $10^{18}$ based on the potential number of data in future Internet. Each ID updates its IPs 100 times per day. According to Cisco forecast, the portion of audio-visual Internet service traffic in the total amount of Internet service traffic keeps increasing to potentially reach around 80% next year, so we assume the 80% memory of each network element is for audio-visual Internet service, and the remaining 20% memory is for other services. The downlink bandwidths are 3 times wider than the uplink bandwidths. The initial value is set to be 8 GB memory and 1.2 Gbps downlink bandwidths with 6.4 GB free memory and 0.4 Gbps uplink bandwidths for Internet service.

For the simulation with satisfying the 5G requirements, a server with 24 processors of 2.10GHz Intel Xeon E5-2620 CPU 32GB RAM and NVIDIA GPU Grid K2 8GB GDDR5 and many computers with two processors of 2.50GHz Intel Core i7-4710MQ CPU 16GB RAM and NVIDIA GEFORCE GTX 760 GPU 2GB GDDR5 are used to train the learning system and constructing the containers. Three-level containers are constructed to simulate typical 5G scenarios for future audio-visual-tactile Internet. We assume that the URLLC scenario requires $T_1$ for less than 1ms ultra-low latency, the eMBB scenario requires $T_2$ for 1~150ms low latency, and the mMTC scenario requires $T_3$ for 150~500ms latency. The URLLC and eMBB scenarios could keep the service continuity in movements with different speeds such as 5km/s, 50km/s, and 500km/s. The physically distributed or collaborative machines could be implemented by nested containers in a logically centralized way to improve the scalability as the size and dynamics of networks increase.

For quantitative evaluation based on audio-visual-tactile multimedia service, 9,581,751 imbalanced samples with partial and conflicting labels from the history are collected as a dataset. There are 9,077,448 benign samples, 328, 811 malware samples and 175,492 uncertain samples from ICN servers, personal computers, devices, sensors, routers, switches, network performance collectors, user-generated data servers, caching servers, and online web sites. Besides, the open-source ISCX2012 [31] could be used as a part of dataset, if the security [32][33] issues are considered in the near future.

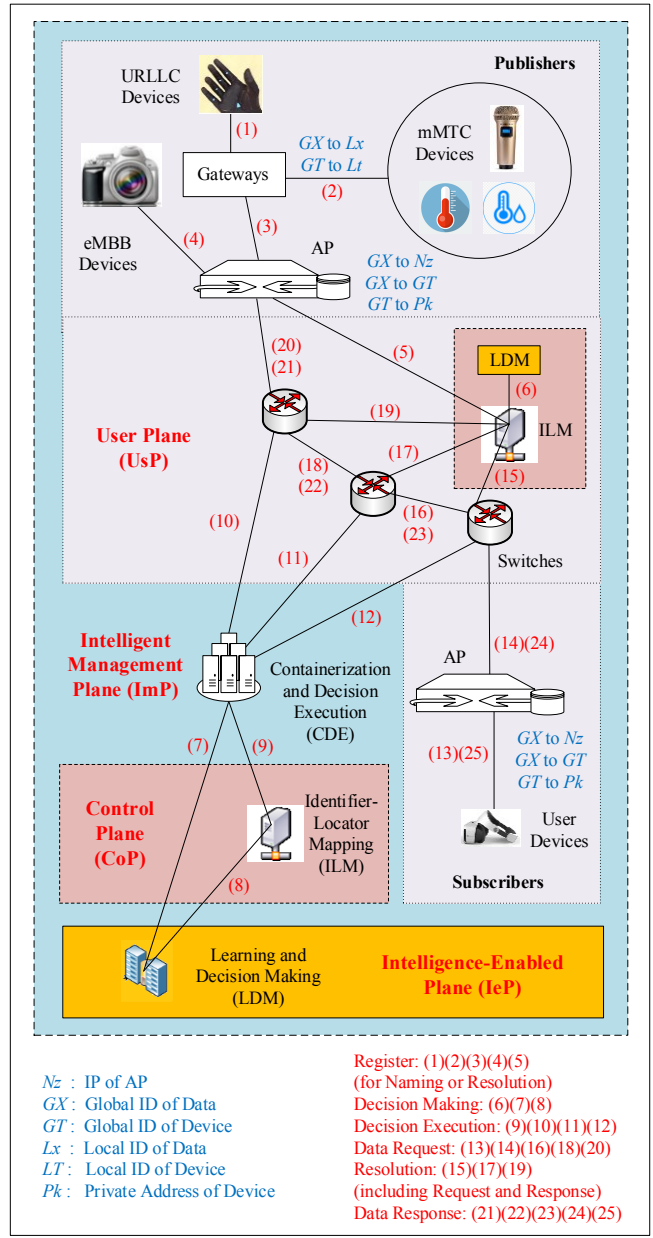

**FIGURE 5.** The evaluation for the scalable performance based on assumption, simulation and proof-of-concept (PoC) implementation using information-centric networking.

The evaluation of SIEN is as shown in FIGURE 5. In containers, a message can be propagated among multiple levels, so that a binding for an object is recorded along the path from one level to another level. For example, a message is propagated from the user plane on the bottom level, to the control plane on the middle levels, and then to the IeP on the top level.

For the publishers and subscribers in the private routing area, if a device is registered in the ILMs with naming service to get a global ID (marked as *GT*) for publishing and subscribing, and the *GT* is mapped to its private address *Pk*. If the device is in local mMTC domain, it has a short name (marked as *Lt*) as its local identifier (LID). The data object, which has its global ID (marked as *GX*), can be registered in the private area to make it accessible for other objects connected to the same access point (AP). If the data object is in the local mMTC domain, it should also has a short name (marked as *Lx*) as its LID. Each AP has its IP (marked as *Nz*), and an indirect binding should map the global data identifier (*GX*) into the device identifier (*GT*). Inside the local mMTC domain, the indirect binding should be from an mMTC data identifier (*Lx*) to an mMTC device identifier (*Lt*).

For the PoC implementation in the public routing area, the data registration is generated from an URLLC device (1), an mMTC device (2), or an eMBB device (3). The registration from the URLLC or mMTC devices would be propagated to the gateway (4) for the local domain of the scenario. Then it would be propagated to the ILM with naming service on the control plane as (5) via an AP to finish the registration. Jingling Phantom 4 Pro unmanned aerial vehicle and Raspberry Pi 3Model B with thermo-hygrostat sensors and cameras for augmented reality can be used as the devices and the gateways. The process of decision making is as shown in (6)(7)(8)(9). The network data would be gathered from the control plane and the ImP to the IeP as (6)(7). The result of decision making would be propagated directly or indirectly from the LDMs on the IeP to the ILMs on the control plane as (8), and then to a set of network elements in multi-level containers on the ImP as (9). The decisions such as prefetching would be executed as (10)(11)(12) for delivering the top-$k$ popularity objects as the local copies for Internet service in the distributed network elements with storage spaces.

When a data request is from the subscribers, it would be propagated to a local AP as (13) and then to a switch as (14). The switch would forward the request to ILMs as (15) so that at least one IP is returned as the result of mapping. Then, the switch keeps forwarding the data request to other switches hop by hop as (16)(18)(20) till reaching the returned IP. Each switch could forward the request to its local ILM as (17)(19) for the ILMs so that the latest IPs are returned as the result of mapping. And then, the requested data could be propagated to the subscribers as shown in (21)(22)(23)(24)(25). The above situation is the worst case. In the better case, the requested data would be prefetched on the forwarding path. The data could be propagated to the subscribers with a part of the path from (14) to (24), so the network traffic would be offloaded.

In order to measure future network traffic performance more accurately in 5G scenarios for audio-visual-tactile multimedia service, Internet traffic offloading (*ITO*) is newly designed as the following.

$$ITO = \frac{\sum_{n=1}^{N}(J_n H_{cn} - \sum_{j=1}^{J_n} H_{nj})V_n}{\sum_{n=1}^{N} J_n H_{cn} V_n}.$$

where $N$ means the total number of requests from users, $V_n$ denotes the volume of the $n$-th requested object, $H_{nj}$ represents the hop count of the $j$-th routing path in the $n$-th request between the users and the in-network object sources, $1 \leq n \leq N$, $1 \leq j \leq J_n$, and $H_{cn}$ is the hop count of the routing without LDM function, prefetching and caching in the $n$-th request between the users and the original object sources. The *ITO* should be more accurate to evaluate the in-network traffic than the hit rate without considering the object volume and the hop count. The volume of objects should be taken into account. The hop count between the users and the sources would be one of the most important factors to impact the overall flow, since the overall in-network traffic offloading of the 1-hop hit between devices is obviously higher than that of the $n$-hop hit between personal computers.

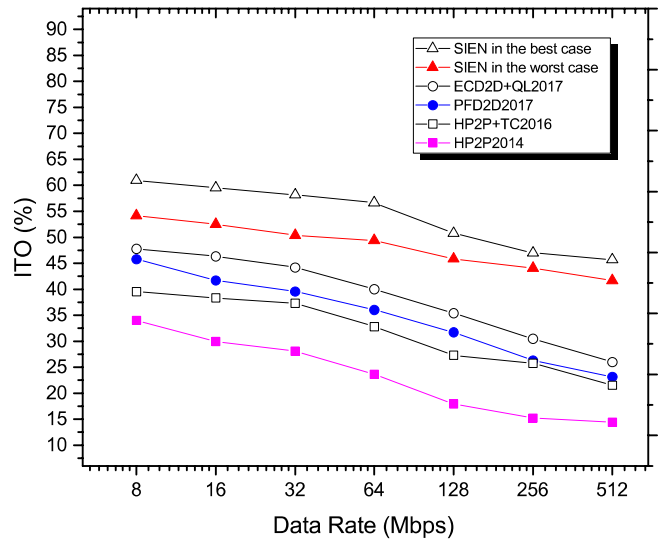

**FIGURE 6.** The performance of eMBB visual traffic offloading with the scalability of the average data rate using the SIEN in comparison to the state-of-the-art techniques [7][8][9][10].

As shown in FIGURE 6, the performance of eMBB visual traffic offloading was evaluated with respect to the scalability of the average data rate from 8 Mbps to 512 Mbps using the SIEN in comparison to the state-of-the-art techniques. The SIEN improved the performance of eMBB visual traffic offloading by maximum 19.74%, average 15.44% and minimum 13.16% in the best case and by maximum 15.71%, average 9.92% and minimum 6.17% in the worst case in comparison to ECD2D+QL2017. The SIEN improved the performance of eMBB visual traffic offloading by maximum 22.56%, average 19.02% and minimum 15.17% in the best case and by maximum 18.53%, average 13.50% and minimum 8.38% in the worst case in comparison to PFD2D2017. The SIEN improved the performance of eMBB visual traffic offloading by maximum

24.21%, average 22.08% and minimum 20.76% in the best case and by maximum 20.18%, average 16.55% and minimum 13.09% in the worst case in comparison to HP2P+TC2016. The SIEN improved the performance of eMBB visual traffic offloading by maximum 33.03%, average 29.71% and minimum 25.05% in the best case and by maximum 28.88%, average 24.19% and minimum 20.20% in the worst case in comparison to HP2P2014.

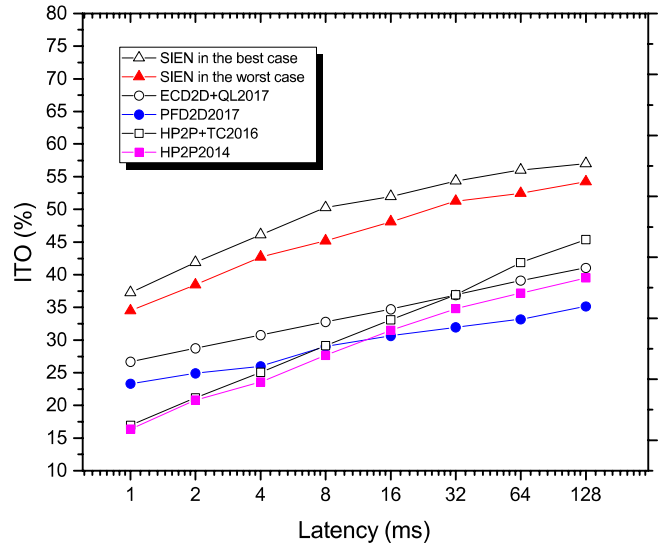

**FIGURE 7.** The performance of URLLC tactile traffic offloading with the scalability of the average latency using the SIEN in comparison to the state-of-the-art techniques [7][8][9][10].

As shown in FIGURE 7, the performance of URLLC tactile traffic offloading was evaluated with respect to the scalability of the average latency from 1 *ms* to 128 *ms* using the SIEN in comparison to the state-of-the-art techniques. The SIEN improved the performance of URLLC tactile traffic offloading by maximum 17.50%, average 14.76% and minimum 8.69% in the best case and by maximum 14.38%, average 11.30% and minimum 5.56% in the worst case in comparison to ECD2D+QL2017. The SIEN improved the performance of URLLC tactile traffic offloading by maximum 22.83%, average 19.32% and minimum 13.15% in the best case and by maximum 19.31%, average 15.86% and minimum 10.02% in the worst case in comparison to PFD2D2017. The SIEN improved the performance of URLLC tactile traffic offloading by maximum 21.13%, average 18.34% and minimum 11.61% in the best case and by maximum 17.63%, average 14.89% and minimum 8.86% in the worst case in comparison to HP2P+TC2016. The SIEN improved the performance of URLLC tactile traffic offloading by maximum 22.61%, average 20.41% and minimum 17.46% in the best case and by maximum 19.14%, average 16.96% and minimum 14.71% in the worst case in comparison to HP2P2014.

As shown in FIGURE 8, the performance of mMTC hygro-thermo-audio traffic offloading was evaluated with respect to the scalability of the average density from 63 kilo objects per km$^2$ to 1049 kilo objects per km$^2$ using the SIEN in comparison to the state-of-the-art techniques. The SIEN improved the performance of mMTC hygro-thermo-audio traffic offloading by maximum 28.41%, average 16.74% and minimum 7.10% in the best case and by maximum 17.45%, average 11.72% and minimum 5.90% in the worst case in comparison to ECD2D+QL2017. The SIEN improved the performance of mMTC hygro-thermo-audio traffic offloading by maximum 32.65%, average 20.86% and minimum 7.80% in the best case and by maximum 21.70%, average 15.85% and minimum 6.60% in the worst case in comparison to PFD2D2017. The SIEN improved the performance of mMTC hygro-thermo-audio traffic offloading by maximum 36.37%, average 25.13% and minimum 8.44% in the best case and by maximum 27.32%, average 20.12% and minimum 7.24% in the worst case in comparison to HP2P+TC2016. The SIEN improved the performance of mMTC hygro-thermo-audio traffic offloading by maximum 46.04%, average 31.97% and minimum 11.47% in the best case and by maximum 36.71%, average 26.96% and minimum 10.27% in the worst case in comparison to HP2P2014.

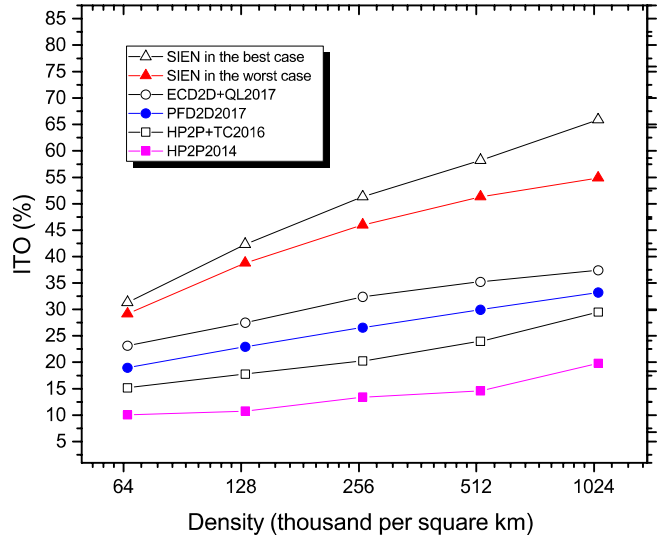

**FIGURE 8.** The performance of mMTC hygro-thermo-audio traffic offloading with the scalability of the average density using the SIEN in comparison to the state-of-the-art techniques [7][8][9][10].

## VII. RELATED WORK

It is urgent for the traffic offloading techniques to satisfy the requirements of future Internet such as 5G. For vehicle-to-vehicle (V2V) scenario, an information-centric networking (ICN) based caching policy was introduced by W. Zhao *et al.* [14] to improve traffic offloading. An ICN-based framework was introduced by G. Chandrasekaran *et al.* [15] for mobile distribution using device-to-device (D2D) for achieving traffic offloading. To exploit ICN chunks by prefetching, a caching scheme was introduced by G. Rossini *et al.* [34] to move solid state drives (SSD) bottleneck from access time to external data rate. By predictive network traffic engineering, a log-normal video-on-demand (VoD) distribution model was proposed by H. Hasegawa *et al.* [35] to show the effectiveness for online streaming service. Since the traffic from online mobile video delivery that provides the resources of VoD services has occupied an increasingly large fraction of traffic, multimedia streaming would be considered as one of the major parts of network traffic overhead. The practices of D2D techniques were summarized by X. Lin *et al.* [36] with identifying challenges and drawing lessons.

The intelligent networks become increasingly important for large-scale cyber-physical distributed systems such as Internet of things. The programmable networking-computing integration was discussed by J. H. Haga *et al.* [22] on using SDN by intelligent solutions. A knowledge plane was proposed by D. D. Clark *et al.* [25] for the Internet. To overcome the shortcomings of the knowledge plane, an inference plane was proposed by J. Strassner *et al.* [26] as a decentralized network overlay for cognitive information processing. By focusing on the reasoning on network management with a knowledge plane, a knowledge graph was introduced by R. Quinn *et al.* [37] to capture data on the networks and the applications. The knowledge plane was used by storing information concerning all layers of the protocol stack for mobile wireless ad hoc networks (MANETs) to improve the network performance [38] and to reduce the energy consumption. By focusing on fault management for multi-domain services, a shared knowledge plane was implemented by A. Castro *et al.* [39] to ensure communication among distributed agents. However, it is still hard to realize scalable intelligent networks by taking full advantage of large-scale data and heterogeneous resources.

## VIII. ANALYSIS AND DISCUSSION

In this section, the validity, the major factors to impact the performance, and the potential researches on SIEN in future Internet would be analyzed and discussed.

For the validity, both the best case and the worst case have been evaluated. According to the statistics of traffic offloading with application to 5G scenarios [40], the overall performance is improved by loading CDM, LDM and ILM with flat names for objects in practice, but the reason why it works is still unclear in theory. The average variance of traffic offloading is 5.52% with respect to the data rate in eMBB scenario. The average variance is 3.40% with respect to the latency in URLLC scenario. The average variance is 5.01% with respect to the density in mMTC scenario. It would be more valid to have real experiments based on a global infrastructure.

The ImP with containerization should be one of major factors to improve the network performance. The CDE is used to realize proximity-based data sharing and traffic localization. It is the foundation to deploy ILMs on the control plane. The ILMs also improve the performance by supporting large-scale dynamic bindings between identifiers and locators. The hierarchical ILM-based scheme makes it possible to share fine-grained data for dense networks with on-demand mobility. The main difference between ILM and DNS is as follows. The ILM is based on flat names in the length of 160 bits, while the

DNS is based on hierarchical IPs. The flat name space, which is $2^{160}$ or near $10^{48}$, is much larger than the IP-based hierarchical name space. The name authenticity can be verified by the self-certifying scheme or the existing certifying schemes such as PKI and block chain to ensure the data integrity.

The IeP with congruity learning should be another major factor to improve the performance. The prefetching should be positive for traffic offloading. The IeP offers a global view to prefetch the named data based on LDM and CDE. The LDM can be used for the distance prediction, the state prediction, and the data classification. The imbalanced, conflicting and partial data can be purified by the LDM. It will be meaningful to research spatiotemporal learning with data purification and imbalanced learning [41].

As shown in FIGURE 9, Akhshabi *et al* [42] pointed out that the internet protocol stack has a hourglass structure, and the current thin waist appears in the IP layer. The current Internet suffers the problem of cross-layer incongruity with many serious issues such as scalability, mobility and security. The higher layer over IP includes protocols, applications, and services. The lower layer under IP includes the infrastructure such as mobile terminals and heterogeneous networks. It is hard for the current IP-based architecture to meet the requirements of future Internet [43] with the development of emerging applications such as audio-visual-tactile multimedia service, ultra-low-latency interaction, and reliable telemedicine. To take differentiated service in audio-visual-tactile Internet as an example, the auditory response time is within 100 millisecond, the visual response time is about 10 millisecond, and the tactile response time is close to a millisecond.

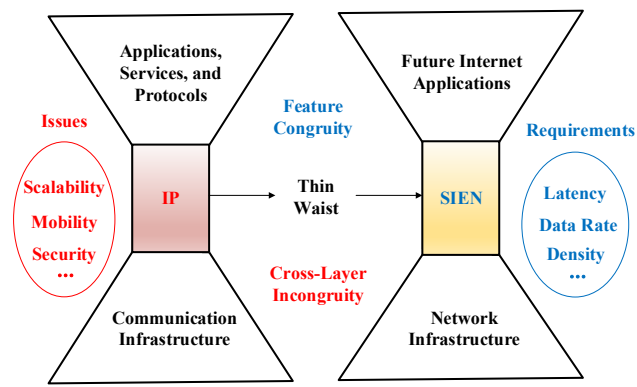

**FIGURE 9.** The SIEN as the thin waist of the hourglass structure to improve cross-layer feature incongruity which impacts many issues and requirements in future Internet.

There are three basic ways to overcome the above issues of IP thin waist. The first one is incremental-patch-first such as Diffserv [44]. The second one is clean-slate-first [45] such as ICN. The third one is scalability-first with IP compatibility. The SIEN with all named objects, subjects and entities would be the scalable thin waist between future Internet applications and network infrastructure to satisfy the increasing requirements such as latency, data rate and density, where an entity is something that exists as itself, a subject is an observing being, and an object is an observed thing. The congruity means the awareness and adaptability of cross-layer features in a hierarchical architecture to connect different layers in a more intelligent way. Since most of applications or users are concerned with the information itself, rather than the network elements or hosts to provide the information, it is an incongruity between information-centric application pattern and host-centric communication pattern. The cross-layer incongruity with pattern, topology and spatio-temporal features leads to the degradation of network performance with a series of specific problems such as link failure, end-to-end delay, interactive instability, routing congestion, etc. The current approaches include delay tolerant network, wireless mesh network, content delivery network [46], ultra-dense network [47], cognitive radio network [48], peer-to-peer, machine-to-machine [49], device-to-device [50], and so on.

In SIEN, each object is named by a globally unique flat name to support data communication and service delivery using the connectionless model by binding the object to the identifier with its locators. Any application can use the identifier to obtain the corresponding service with at least one locator such as IP for routing, forwarding and mobility management. The data can be exchanged between internal and public networks by gateways with a mapping between local and global identifiers. It allows the provisioning of diverse objects such as devices, data and services. A fixed-length globally unique identifier would be allocated to each object with self-verification, and the data integrity could be achieved.

By considering the scalability of mobile users, network elements and data objects, the LDM on the IeP can be used as one of network functions in future Internet to improve performance by building self-configuring, self-managing and self-optimizing networks. The related network functions in 5G may include user equipment, radio access network or access network, user plane function, data network, etc. The data rate, latency and connection density are three of 5G basic requirements. For the 5G eMBB scenario, the user data rate would be from 0.1Gbps to 1Gbps, or even 20Gbps, which should be 100 times or even higher user data rate than 4G. For the 5G URLLC scenario, the latency would be 1ms or even lower, which should be at least

5 times reduced latency than 4G. For the 5G mMTC scenario, the connection density would be from $10^3$ to $10^6$ devices or more objects per $km^2$, which should be from 10 to 100 times or even higher connection number than 4G. In fact, the international telecommunication union has identified fifteen ICN-related standardization gaps. This paper covers many gaps such as E.1, E.8, E.9 and E.13. The related works have proved ICN with exploiting the communications between proximity-based objects such as D2D and P2P improved performance such as traffic offloading, spectrum utilization, overall throughput, energy consumption, etc.

As the user equipment (UE), Internet service providers (ISPs) and Internet content providers (ICPs) increases rapidly, the current Internet scales with heterogeneous interactive network elements accessing to it. The traditional networking uses a single ISP node for a large number of users, even if the ISP node has insufficient capacity. In the current networks, different autonomous systems (ASs) are managed by different ISPs. As a result, each ISP upgrade the infrastructure within its own ASs due to the economic benefits. That leads to many inter-domain issues such as latency, bandwidth, and service quality. The inter-domain routing is based on the BGPs which select only one path at a time. Any failure on the path may lead to an interruption or a network shock. Especially for a larger-scale network, the interruptions or shocks become frequent, so it leads to a big issue of scalability.

The SIEN is designed to intelligently access any a named object as a service offered by the network. It is supposed to be more resilient to deal with the failures such as the inter-domain routing shock or churn. It is supposed to improve QoX for users across ISPs or ICPs with multiple ASs. It is supposed to improve the capacity of multi-path Internet service by facing the challenge with the rapidly increasing number of the UEs such as IoT devices. It is supposed to utilize large-scale network resources and provide any Internet service for any user at anytime and anywhere.

This paper showed the feasibility for SIEN to deliver an open data service in 5G scenarios for audio-visual-tactile Internet. The SIEN supports identifier-locator separation, and it offers the flexibility for diverse objects to move among multiple ISPs, ICPs, or ASs by dynamical binding between identifiers and locators. Because each identifier is unique and persistent, a conversation can be established by the identifier and routed by its locator. The level of containers depends on the spatio-temporal range such as latency. The different levels of ILMs support the object registration and resolution from UEs with different moving speeds. The requests of registration and resolution would be forwarded to one of the most appropriate ILMs to get high-performance response regardless of the location and the moving speed of an object. Thus it is beneficial to service continuity by supporting both subscriber and publisher mobility. The IPv4 or IPv6 can be used as locator to achieve the compatibility with IP-based infrastructure.

As shown in FIGURE 10, the SIEN focuses on a new research topic on cross-layer feature congruity in future Internet. In this young topic, many problems are still open, immature or undefined, and we do not know what the mechanisms really are and how to build them in details. First, based on the best-so-far understanding for feature congruity, the SIEN can be modelled by the ImP with CDEs, the IeP with LDMs, the CoP with ILMs and the UsP with agents/drivers. Second, the proof-of-concept system for future audio-visual-tactile Internet is implemented based on the technical details of SIEN. Third, the performance is evaluated to get feedback for improving the technical details. Fourth, the experimental evidences are collected to get better understanding for cross-layer feature congruity as a research loop of continuous improvement.

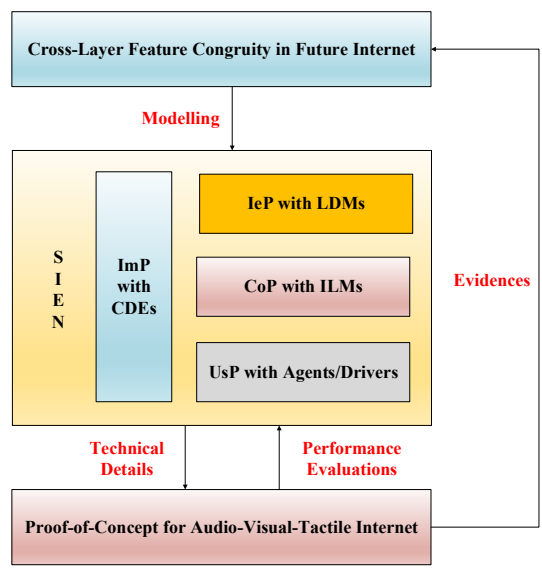

**FIGURE 10.** A research loop of continuous improvement to get better understanding for cross-layer feature congruity in future networks such as audio-visual-tactile Internet.

## IX. CONCLUSION

The scalable intelligence-enabled networking (SIEN) was proposed with a mix of assumption, simulation and proof-of-concept implementation based on ICN in typical 5G scenarios for future audio-visual-tactile Internet, and a technique for Internet traffic offloading was contributed. The SIEN included an intelligence-enabled plane (IeP) on the top level, an intelligent management plane (ImP) with more than three levels, a control plane on the middle levels, and a user plane on the bottom level. On the IeP, a congruity learning with generalization with personalization was contributed for learning and decisions making (LDM). On the ImP, a weighted graph algorithm was contributed to organize network elements or resource partitions for containerization and decision execution (CDE). On the control plane, a CDE-based scheme was contributed for identifier-locator mapping (ILM). On the user plane, the registrations, requests and data would be forwarded to test the traffic performance. The evaluation showed that the SIEN outperformed four state-of-the-art techniques of traffic offloading, as the change of the data rate in eMBB scenario, the latency in URLLC scenario and the density in mMTC scenario.

In future, it would be meaningful for SIEN to study locality-sensitive CDE, spatio-temporal LDM, enhanced ILM, and on-site real-time mobile edge service. It would be advantageous for SIEN to improve multimedia, multipath, multi-point, multi-protocol, multi-tenant, multi-level, cross-layer and inter-domain problems. It would be beneficial to solve more issues such as security, compatibility, latency, mobility, routing, caching, load balance, congestion control, resource merger/segmentation, manageability, deployability, scheduling, resiliency, and so on.

## REFERENCES

[1] R. Fontugne, P. Abry, K. Fukuda, D. Veitch, K. Cho, P. Borgnat, and H. Wendt, "Scaling in Internet Traffic: A 14 Year and 3 Day Longitudinal Study, With Multiscale Analyses and Random Projections," IEEE/ACM Transactions on Networking, vol. 25, no. 4, pp. 2152 - 2165, 2017.

[2] L. Dong, H. Zhao, Y. Chen, D. Chen, T. Wang, L. Lu, B. Zhang, L. Hu, L. Gu, B. Li, H. Yang, H. Shen, T. Tian, Z. Luo, and K. Wei, "Introduction on IMT-2020 (5G) Trials in China," IEEE Journal on Selected Areas in Communications, vol 35, no. 8, pp 1849 - 1866, 2017.

[3] X. Yang, Z. Huang, B. Han, S. Zhang, C. Wen, F. Gao, S. Jin, "RaPro: A Novel 5G Rapid Prototyping System Architecture," IEEE Wireless Communications Letters, vol 6, issue 3, pp 362 - 365, 2017.

[4] T. Taleb, K. Samdanis, B. Mada, H. Flinck, S. Dutta, D. Sabella, "On Multi-Access Edge Computing: A Survey of the Emerging 5G Network Edge Cloud Architecture and Orchestration," IEEE Communications Surveys & Tutorials, vol 19, issue 3, pp 1657 - 1681, 2017.

[5] Y. Mehmood, N. Haider, M. Imran, A. Timm-Giel, M. Guizani, "M2M Communications in 5G: State-of-the-Art Architecture, Recent Advances, and Research Challenges," IEEE Communications Magazine, vol 55, issue 9, pp 194 - 201, 2017.

[6] J. Pilz, M. Mehlhose, T. Wirth, D. Wieruch, B. Holfeld, and T. Haustein, "A Tactile Internet demonstration: 1ms ultra low delay for wireless communications towards 5G," in 2016 IEEE Conference on Computer Communications Workshops (INFOCOM WKSHPS), pp. 862 - 863, 2016.

[7] W. Wang, R. Lan, J. Gu, A. Huang, H. Shan, and Z. Zhang, "Edge Caching at Base Stations With Device-to-Device Offloading," IEEE Access, vol 5, pp 6399 - 6410, 2017.

[8] G. Park, W. Kim, S. Jeong, and H. Song, "Smart Base Station-Assisted Partial-Flow Device-to-Device Offloading System for Video Streaming Services," IEEE Transactions on Mobile Computing, vol 16, issue 9, pp 2639 - 2655, 2017.

[9] C. Tseng, G. Huang, and T. Liu, "P2P traffic classification using clustering technology," 2016 IEEE/SICE International Symposium on System Integration (SII), pp 174 - 179, 2016.

[10] H. Hoang-Van, T. Miyoshi, and O. Fourmaux, "A hierarchical P2P traffic localization method with bandwidth limitation," The IEEE International Conference on Communications (ICC), pp 3136 - 3141, 2014.

[11] D. Xu, Y. Li, J. Li, M. Ahmed, and Pan Hui, "Joint Topology Control and Resource Allocation for Network Coding Enabled D2D Traffic Offloading," IEEE Access, vol 5, pp 22916 - 22926, 2017.

[12] P. Casas, A. D'Alconzo, P. Fiadino, A. Bär, A. Finamore, and T. Zseby, "When YouTube Does not Work: Analysis of QoE-Relevant Degradation in Google CDN Traffic," IEEE Transactions on Network and Service Management, vol. 11, issue 4, pp 441 - 457, 2014.

[13] M. Hu, Z. Zhong, M. Ni, Z. Wang, W. Xie, and X. Qiao, "Integrity-Oriented Content Offloading in Vehicular Sensor Network," IEEE Access, vol 5, pp 4140 - 4153, 2017.

[14] W. Zhao, Y. Qin, D. Gao, C. H. Foh, and H. Chao, "An Efficient Cache Strategy in Information Centric Networking Vehicle-to-Vehicle Scenario," IEEE Access, vol. 5, pp. 12657 - 12667, 2017.

[15] G. Chandrasekaran, N. Wang, and R. Tafazolli, "Caching on the move: Towards D2D-based information centric networking for mobile content distribution" The IEEE 40th Conference on Local Computer Networks (LCN), pp. 312 - 320, 2015.

[16] A. Hussein, I. Elhajj, A. Chehab, A. Kayssi, "SDN VANETs in 5G: An architecture for resilient security services," The Fourth International Conference on Software Defined Systems (SDS), pp 67 - 74, 2017.

[17] G. Xylomenos et al., A survey of information-centric networking research, IEEE Commun. Surveys Tuts., vol. 16, no. 2, pp. 1024–1049, 2nd Quart., 2014.

[18] B. Ahlgren, C. Dannewitz, C. Imbrenda, D. Kutscher, and B. Ohlman, "A survey of information-centric networking," IEEE Commun. Mag., vol. 49, No. 7, pp. 26–36, Jul. 2012.

[19] M. F. Bari, S. R. Chowdhury, R. Ahmed, R. Boutaba, and B. Mathieu, "A survey of naming and routing in information-centric networks," IEEE Commun. Mag., vol. 49, No. 12, pp. 44–53, Dec. 2012.

[20] M. D'Ambrosio, C. Dannewitz, H. Karl, and V. Vercellone, "MDHT: A hierarchical name resolution service for information-centric networks," in Proc. of ACM SIGCOMM Workshop ICN, pp. 7–12., 2011.

[21] D. Lagutin, K. Visala, and S. Tarkoma, "Publish/subscribe for Internet: PSIRP perspective," in Towards the Future Internet, vol. 4. Amsterdam, The Netherlands: IOS Press, pp. 75–84, 2010.

[22] J. H. Haga, C. Fernandez, A. Takefusa, T. Ikeda, J. Tanaka, B. Belter, and T. Kudoh, "Building Intelligent Future Internet Infrastructures: FELIX for Federating Software-Defined Networking Experimental Networks in Europe and Japan," IEEE Systems, Man, and Cybernetics Magazine, vol. 3, no. 4, pp. 35 - 42, 2017.

[23] M. Zorzi, A. Zanella, A. Testolin, M. F. Grazia, and Ma. Zorzi, "Cognition-Based Networks: A New Perspective on Network Optimization Using Learning and Distributed Intelligence," IEEE Access, vol. 3, pp. 1512 - 1530, 2015.

[24] P. Barlet-Ros, and G. Iannaccone, "Predictive resource management of multiple monitoring applications," IEEE/ACM Transactions on Networking, vol.19, no.3, pp 788-801, June 2011.
[25] D. D. Clark, C. Partridge, J. C. Ramming, and J. T. Wroclawski, "A knowledge plane for the Internet," Proceedings of the 2003 Conference on applications, technologies, architectures, and protocols for computer communications, ACM SIGCOMM, pp 3–10, 2003.
[26] J. Strassner, M. OFoghlu, W. Donnelly, and N. Agoulmine, "Beyond the Knowledge Plane: An Inference Plane to Support the Next Generation Internet," The First International Global Information Infrastructure Symposium, pp. 112 - 119, 2007.
[27] X. Wang, X. Li, and V. C. M. Leung, "Artificial Intelligence-Based Techniques for Emerging Heterogeneous Network: State of the Arts, Opportunities, and Challenges," IEEE Access, vol. 3, pp. 1379 - 1391, 2015.
[28] C. Wu, M. Gales, A. Ragni, P. Karanasou, and K. Sim, "Improving Interpretability and Regularization in Deep Learning," IEEE/ACM Transactions on Audio, Speech, and Language Processing, vol. 26, issue 2, pp 256 - 265, 2018.
[29] S. Biswas, and P. Milanfar, "One Shot Detection with Laplacian Object and Fast Matrix Cosine Similarity," IEEE Transactions on Pattern Analysis and Machine Intelligence, vol. 38, issue 3, pp 546 - 562, 2016.
[30] A. Nguyen, V. Reju, A. Khong, and I. Soon, "Learning complex-valued latent filters with absolute cosine similarity," The IEEE International Conference on Acoustics, Speech and Signal Processing (ICASSP), pp 2412 - 2416, 2017.
[31] KDN2016 Datasets. Accessed: Feb. 2017. [Online]. Available: http://knowledgedefinednetworking.org/
[32] N. Hubballi, and V. Suryanarayanan, "False alarm minimization techniques in signature-based intrusion detection systems: A survey", Computer Communications, vol. 49, no. 1, pp. 1-17, 2014.
[33] T. Ma, F. Wang, J. Cheng, Y. Yu, X. Chen, "A hybrid spectral clustering and deep neural network ensemble algorithm for intrusion detection in sensor networks", Sensors, vol. 16, no. 10, pp. 1701, 2016.
[34] G. Rossini, D. Rossi, M. Garetto, and E. Leonardi, "Multi-terabyte and multi-gbps information centric routers," The 33rd IEEE International Conference on Computer Communications (INFOCOM), pp. 181–189, 2014.
[35] H. Hasegawa, S. Kouno, A. Shiozu, M. Sasaki, and S. Shimogawa, "Predictive network traffic engineering for streaming video service," The IFIP/IEEE International Symposium on Integrated Network Management, pp. 788 - 791, 2013.
[36] X. Lin, J. G. Andrews, A. Ghosh, and R. Ratasuk, "An Overview on 3GPP Device-to-Device Proximity Services," IEEE Communications Magazine, vol 52, no. 4, pp 40 - 48, 2013.
[37] R. Quinn, J. Kunz, A. Syed, J. Breen, S. Kasera, R. Ricci, and J. V. Merwe, "KnowNet: Towards a knowledge plane for enterprise network management," IEEE/IFIP Network Operations and Management Symposium, pp 249 - 256, 2016.
[38] D. R. Macedo, A. L. dos Santos, G. Pujolle, and J. M. S. Nogueira, "Optimizing wireless ad hoc communication resources with a knowledge plane," The 9th IFIP International Conference on Mobile Wireless Communications Networks, pp. 86 - 90, 2007.
[39] A. Castro, B. Fuentes, J. A. Lozano, B. Costales, and V. Villagrá, "Multi-domain fault management architecture based on a shared ontology-based knowledge plane," IEEE Communications Magazine, pp 493 - 498, 2010.
[40] E. Pateromichelakis, J. Gebert, T. Mach, J. Belschner, W. Guo, N. Kuruvatti, V. Venkatasubramanian, and C. Kilinc, "Service-Tailored User-Plane Design Framework and Architecture Considerations in 5G Radio Access Networks," IEEE Access, vol 5, pp 17089 - 17105, 2017.
[41] H. He and E. A. Garcia, "Learning from Imbalanced Data," in IEEE Transactions on Knowledge and Data Engineering, vol. 21, no. 9, pp. 1263-1284, Sept. 2009.
[42] P. Calyam, M. Sridharan, Y. Xu, K. Zhu, A. Berryman, R. Patali, and A. Venkataraman, "Enabling performance intelligence for application adaptation in the Future Internet," Journal of Communications and Networks, vol. 13, issue 6, pp. 591 - 601, 2011.
[43] S. Akhshabi, and C. Dovrolis, "The evolution of layered protocol stacks leads to an hourglass-shaped architecture," in ACM SIGCOMM Computer Communication Review, vol. 41, issue 4, pp. 206-217, 2011.
[44] Diffserv: https://datatracker.ietf.org/wg/diffserv/.
[45] Clean Slate: https://cleanslate.stanford.edu.
[46] J. Liu, Q. Yang, and G. Simon, "Congestion Avoidance and Load Balancing in Content Placement and Request Redirection for Mobile CDN," IEEE/ACM Transactions on Networking, vol. PP, issue 99 (Early Access Articles), 2018.
[47] Q. Liu, G. Chuai, W. Gao, and K. Zhang, "Load-aware user-centric virtual cell design in Ultra-Dense Network," 2017 IEEE Conference on Computer Communications Workshops (INFOCOM WKSHPS), pp. 619 - 624, 2017.
[48] A. Omer, M. Hassan, and M. El-Tarhuni, "Window-based adaptive technique for real-time streaming of scalable video over cognitive radio networks," IET Communications, vol. 11, issue 17, pp. 2643 - 2649, 2017.
[49] W. Li, Q. Du, L. Liu, P. Ren, Y. Wang, and L. Sun, "Dynamic Allocation of RACH Resource for Clustered M2M Communications in LTE Networks," 2015 International Conference on Identification, Information, and Knowledge in the Internet of Things (IIKI), pp. 140 - 145, 2015.
[50] X. Wang, Y. Zhang, V. Leung, N. Guizani, and T. Jiang, "D2D Big Data: Content Deliveries over Wireless Device-to-Device Sharing in Large-Scale Mobile Networks," IEEE Wireless Communications, vol. 25, issue 1, pp. 32 - 38, 2018.